\documentclass[prb,twocolumn,showpacs,amsmath,amssymb,eqsecnum,superscriptaddress]{revtex4}

\usepackage{graphicx}
\usepackage{dcolumn}
\usepackage{bm}
\begin{document}

\newcommand{\lin}{l_{\textrm{in}}}

\newcommand{\me}{m_{\textrm{e}}}

\newcommand{\tperp}{t_{\perp}}

\newcommand{\Bperp}{B_{\perp}}

\newcommand{\dagg}{^{\dagger}}

\newcommand{\phdagg}{^{\phantom{\dagger}}}

\newcommand{\lB}{l_{\textrm{B}}}

\newcommand{\lel}{l_{\textrm{el}}}

\newcommand{\LT}{L_{\textrm{T}}}

\newcommand{\vF}{v_{\textrm{F}}}

\newcommand{\kB}{k_{\textrm{B}}}

\newcommand{\U}{\mathcal{U}}

\newcommand{\Uzero}{\mathcal{U}_{\,0}}

\newcommand{\Ulin}{\mathcal{U}_{\,\textrm{lin}}}

\newcommand{\V}{\mathcal{V}}

\newcommand{\Vzero}{\mathcal{V}_{\,0}}

\newcommand{\Vlin}{\mathcal{V}_{\,\textrm{lin}}}

\newcommand{\modsq}[1]{\vert #1 \vert^2}

\newcommand{\Tr}{\textrm{Tr}}

\newcommand{\hberg}{^{(H)}}

\newcommand{\Ham}{\mathcal{H}}

\newcommand{\order}[1]{\mathcal{O}(#1)}

\newcommand{\overlap}[2]{\langle #1\vert #2\rangle}

\newcommand{\ket}[1]{\vert #1\rangle}

\newcommand{\bra}[1]{\langle #1\vert}

\newcommand{\pages}[1]{ \emph{(#1 pages)}}

\newcommand{\mb}[1]{\mathbf{#1}}

\newcommand{\subtxt}[1]{_{\textrm{#1}}}

\newcommand{\bmat}{\begin{displaymath}}

\newcommand{\emat}{\end{displaymath}}

\newcommand{\bit}{\begin{itemize}}

\newcommand{\eit}{\end{itemize}}

\newcommand{\beq}{\begin{equation}}

\newcommand{\eeq}{\end{equation}}

\newcommand{\bspl}{\begin{split}}

\newcommand{\espl}{\end{split}}

\newcommand{\twomat}[4]{\ensuremath{\left( \begin{array}{rr} #1 &
        #2\\#3 & #4\end{array}\right)}}

\newcommand{\threemat}[9]{\ensuremath{\left( \begin{array}{rrr} #1 &
        #2 & #3\\#4 & #5 & #6\\#7 & #8 & #9 \end{array} \right)}}

\newcommand{\twomatc}[4]{\ensuremath{\left( \begin{array}{cc} #1 &
        #2\\#3 & #4\end{array}\right)}}

\newcommand{\threematc}[9]{\ensuremath{\left( \begin{array}{ccc} #1 &
        #2 & #3\\#4 & #5 & #6\\#7 & #8 & #9 \end{array} \right)}}

\newcommand{\twovec}[2]{\ensuremath{\left( \begin{array}{r} #1 \\
        #2\end{array}\right)}}

\newcommand{\threevec}[3]{\ensuremath{\left(\begin{array}{r}
        #1\\#2\\#3\end{array}\right)}}

\newcommand{\augmat}[8]{\ensuremath{\left( \begin{array}{rr|rr}
        #1&#2&#3&#4\\#5&#6&#7&#8 \end{array}\right)}}

\preprint{APS/123-QED}

\title{Transport between edge states in multilayer integer 
quantum Hall systems:\\
exact treatment of Coulomb interactions and disorder}
\author{J. W. Tomlinson} 
\affiliation{Theoretical Physics, University of Oxford, 1 Keble
Road, OX1 3NP, United Kingdom.}
\author{J.-S. Caux} 
\affiliation{Institute for Theoretical Physics, University of
Amsterdam, Valckenierstraat 65, 1018 XE Amsterdam, The
Netherlands.}
\author{J. T. Chalker} 
\affiliation{Theoretical Physics, University of Oxford, 1 Keble
Road, OX1 3NP, United Kingdom.}

\date{\today}

\begin{abstract}

A set of stacked two-dimensional electron systems in a perpendicular magnetic field
exhibits a three-dimensional version of the quantum Hall effect if interlayer
tunneling is not too strong. When such a sample is in a quantum Hall
plateau, the edge states of each layer combine to form a chiral
metal at the sample surface. We study the interplay of interactions
and disorder in transport properties of the chiral metal,
in the regime of weak interlayer tunneling. Our starting point is
a system without interlayer tunneling, in which the only excitations 
are harmonic collective modes: surface magnetoplasmons.
Using bosonization and working perturbatively in the interlayer tunneling amplitude,
we express transport properties in terms of the spectrum for
these collective modes, treating electron-electron interactions
and impurity scattering exactly. 
We calculte the conductivity as a function of temperature,
finding that it increases with increasing temperature
as observed in recent experiments. We also calculate the 
autocorrelation function of mesoscopic conductance fluctuations
induced by changes in a magnetic field component perpendicular
to the sample surface, and its dependence
on temperature. We show that conductance fluctuations are characterised by
a dephasing length that varies inversely with temperature.

\end{abstract}




\maketitle

\section{Introduction}\label{sec:intro}

Multilayer quantum Hall systems offer a setting in which
to study the influence of electron-electron interactions
and impurity scattering on tunneling between
quantum Hall edge states. Specifically, consider
a layered conductor in a magnetic field that is perpendicular
to the layers, with the field strength chosen so that a 
single layer in isolation would have quantised Hall conductance.
Then, if interlayer tunneling is not too strong, the multilayer
system exhibits a three-dimensional version of the
quantum Hall effect and the bulk is insulating at low temperatures.
Under these conditions, edge states are present in each layer at the
sample surface and are coupled by interlayer tunneling to form a surface phase,
which is a chiral, two-dimensional metal.\cite{Dohmen,Balents/Fisher}
The contribution of this surface phase to the interlayer electron 
transport properties of such systems has been isolated in experiments
on semiconductor multilayers,\cite{UCSB1}
and is dominant if samples are sufficiently small and cold.

The consequences of impurity scattering for transport in the chiral metal
have been discussed extensively from a theoretical viewpoint 
\cite{Dohmen,Balents/Fisher,Mathur,Gruzberg,Cho,Plerou,Sondhi,Betouras}   
and have been probed experimentally in several 
ways.\cite{UCSB1,UCSB0,MULTILAYER,UCSB9,Kuraguchi/Osada,UCSB5,UCSB7,UCSB2,UCSB11,UCSB10,UCSB12}
Crucially, the chiral motion of electrons along the layer edges
means that localisation is suppressed.\cite{Dohmen,Balents/Fisher}
As a result, the surface conductivity in the interlayer direction
has a low-temperature limit that is non-zero, even though its
measured value may be much smaller than $e^2/h$.\cite{UCSB1,MULTILAYER,Kuraguchi/Osada}
Separately, theoretical discussions
of conductance fluctations\cite{Mathur,Gruzberg,Cho,Plerou,Betouras}
have examined both their dependence on geometry in fully
phase-coherent samples, and their dependence on the inelastic scattering length
when this is smaller than sample size.
Observations of reproducible mesoscopic conductance
fluctuations,\cite{UCSB9,UCSB12}
induced by small changes of magnetic field
within a quantum Hall plateau, demonstrate that interlayer
hopping is quantum-mechanically coherent and also provide a 
way to determine the inelastic scattering length. 
In addition, magnetoresistance in response to a field component
perpendicular to the sample surface has been proposed\cite{Sondhi}
and used\cite{UCSB5,UCSB7,UCSB11} as a method for measuring the
elastic scattering length.

In contrast to these studies of disorder effects, 
past theoretical work on effects due to electron-electron interactions
in the chiral metal has been limited.  There have 
been discussions, first, of the
temperature dependence of the inelastic 
scattering length\cite{Balents/Fisher,Betouras}
and, second, of the fact that there is no zero-bias 
anomaly in the tunneling density
of states (or any related contribution to the conductivity), because
of ballistic motion of charge in the in-layer 
direction.\cite{Balents/Fisher,Betouras}

Against this background, recent experiments finding a significant temperature
dependence to the surface conductivity\cite{UCSB2,UCSB10} are striking as
likely indications of interaction effects, and provide one of the motivations
for the work we present here. In particular, the fact that conductivity
is observed to {\it increase} with increasing temperature presents a puzzle for theory.
Some straightforward potential explanations are specifically excluded
by the experimental design: large ratios of sample perimeter to cross-sectional area
ensure that surface states make the dominant contribution to the measured
conductance; and sample perimeters much longer than the inelastic scattering length 
ensure that weak localisation effects are absent. For samples 
studied in Ref.~\onlinecite{UCSB10}, the measured conductivity $\sigma(T)$
increases by about $7\%$ in the temperature range from $50$mK to $300$mK,
implying a temperature scale of 
$\sigma(T)\cdot [d\sigma(T)/dT]^{-1} \sim 4$K,
which is similar to that for other 
interaction effects in quantum Hall systems 

In this paper we study interactions and disorder in the
chiral metal, working in the experimentally-relevant limit
of weak interlayer tunneling. Treating tunneling perturbatively, 
Coulomb interactions and impurity scattering can be handled
exactly by means of a straighforward application of bosonization.
We calculate the full temperature dependence of the conductivity.
We also study conductance fluctuations induced by magnetic field changes,
obtaining their autocorrelation function and its dependence on
temperature. Making appropriate parameter choices, our results
for both quantities are consistent with
experimental findings. A short account of this work
has been presented previously, in Ref.~\onlinecite{PRL}.

Our work differs from most of the extensive literature on 
tunneling between quantum Hall edges states in two important ways.
First, while much previous work has been concerned with edge
states of fractional quantum Hall systems,\cite{Wen1,Wen2,Wen3,EdgeReview} 
including multilayer 
samples,\cite{Naud1,Naud2}
our focus is on the integer quantum Hall effect. 
Second, whereas most past work (with some exceptions: see
Refs.~\onlinecite{Moon,Zulicke,Oreg,Pryadko}) has been restricted to systems
with only short-range interactions, we find that the long-range
nature of Coulomb interactions, which we treat in full, is 
central for the results we obtain.

The remainder of this paper is organised as follows. We develop a
model for the chiral metal in Sec.~\ref{sec:model} and show how
bosonization can be used to give an exact description of the
collective excitations.  Sec.~\ref{sec:cond} contains calculations
of the temperature dependence of the conductivity. We study
conductance fluctuations in Sec.~\ref{sec:condflucs}, and discuss
our results in Sec.~\ref{sec:discussion}.

\section{Modelling the chiral metal}\label{sec:model}

In this section we summarise the physical ingredients
that are important for modelling transport between edge
states in multilayer conductors and set out the lengthscales
that characterise the system. We introduce a Hamiltionian
in terms of fermionic operators for edge electrons. We
bosonize this Hamiltonian, obtaining a result which
is quadratic in boson operators if interlayer tunneling
is omitted. Finally, we express the two-electron correlation function
that is central to transport calculations in terms of
boson correlators.

\subsection{Ingredients, lengthscales, and parameters}\label{ssec:model:ingreds}

\begin{figure}
\begin{center}
\includegraphics[width=0.40\textwidth]{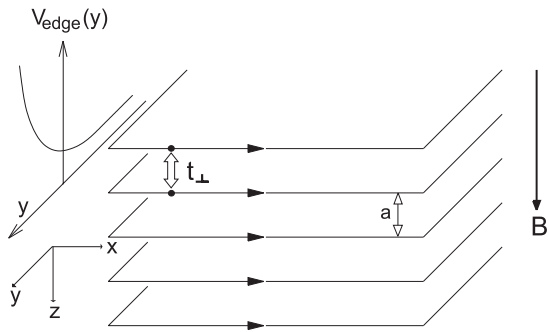}
\caption{\label{fig:stack} A multilayer conductor, showing the orientation of 
axes in our coordinate system, with edge states propagating in the $x$-direction.
The form of the confining potential $V_{\rm edge}(y)$ is illustrated top left.
Interlayer tunneling amplitude and spacing are denoted by $t_\perp$ and $a$,
respectively. }
\end{center}
\end{figure}

A multilayer conductor is illustrated in Fig~\ref{fig:stack}.
We use coordinates with the $x$-axis parallel to the
layer edges, and treat a sample of $N$ layers with layer index $n$
and layer spacing $a$. Consider the system in the presence of a 
perpendicular magnetic field of strength $B$, with the chemical
potential lying between the lowest and first excited Landau levels.
In the bulk of the sample single particle states at energies 
close to the chemical potential are localised by disorder.
At the sample surface in this energy range, edge states
propagate in the confining potential $V_{\rm edge}(y)$ at a velocity $v$.
Interactions modify the confining potential and the edge velocity:
we denote by $v_{\rm F}$ the velocity allowing for Hartree contributions.
Edge states have a width $w$ in the $y$-direction, which is set by the magnetic length
$l_{\rm B}$ in a clean sample, and by the bulk localisation length $\xi$
in the presence of impurities. We use a one-dimensional decription
of the edge state in each layer, projected onto the $x$-coordinate in the standard
way.

Out theoretical treatment takes account only of one edge state 
in each layer and is therefore appropriate for a
system in which electrons are spin polarised.
In fact, some of the experiments we refer to,
including those on the temperature-dependence of conductivity,\cite{UCSB10}
are for systems with Landau level filling factor per layer of $\nu=2$.
It is appropriate to apply our theory
to these systems provided electrons with opposite spin directions
contribute additively and incoherently to the
conductivity. 

The system of edge states can be characterised using three lengthscales.
First, impurities, which generate only forward scattering 
with a phase shift, result in
an elastic mean free path $l_{\rm el}$, the distance over which a phase shift
of order $2\pi$ is accumulated.
Second, temperature $T$ in combination
with the velocity $v_{\rm F}$ can be expressed in terms of the thermal 
length $L_{\rm T} = \hbar v_{\rm F}/k_{\rm B}T$. Third, interlayer tunneling
with amplitude $t_\perp$ can be parameterised
by the characteristic distance $l_\perp$ through which electrons move
in the chiral direction between tunneling events. The value of
$l_\perp$ can be expressed in terms of the interlayer diffusion constant
$D$: since, for small $t_\perp$, interlayer hops are of length $a$ and occur at
a rate $v_{\rm F}/l_\perp$, one has $l_\perp = a^2 v_{\rm F} / D$.
In turn, this can be expressed in terms of the conductivity,
using the Einstein relation and the fact that the
density of states is $n=1/2\pi a\hbar v_{\rm F}$, giving 
$l_\perp =  a (e^2/2\pi \hbar\sigma)$.\cite{Betouras}

Parameter values for the experiments of Refs.~\onlinecite{UCSB1}, \onlinecite{UCSB11}
and \onlinecite{UCSB10} are as follows. Samples consist of $N \sim 50$ -- $100$
layers with spacing $a=30$nm. The mean free path is estimated \cite{UCSB11} to be
$l_{\rm el} \sim 30$nm. An upper bound on $v_{\rm F}$, reached in samples
with a steep confining potential is $v_{\rm F}\sim\omega\subtxt{C}l_{\rm B}$,
where $\omega\subtxt{C}$ is the cyclotron frequency.
It has the value
$\omega\subtxt{C}\lB=1.7\times10^5\textrm{ms}^{-1}$
in GaAs at 6.75 T.
With this value, $L_{\rm T}\sim 10\mu$m at $T=100$mK.
Finally, for a surface conductivity of $\sigma = 1.3 \times 10^{-3}e^2/2\pi \hbar$ 
(which lies within
the observed range at $\nu=2$), $l_\perp = 40\mu$m. We are therefore concerned with the
regime $\lel\ll\LT\ll l_\perp$, and this motivates our approach,
based on a perturbative treatment of tunneling.

\subsection{Fermionic Hamiltonian}\label{ssec:model:fermi}

Our model Hamiltonian,
$\Ham=\Ham_0+\Ham_{\textrm{dis}}+\Ham_{\textrm{hop}}+\Ham_{\textrm{int}}$,
has single-particle terms $\Ham_0$, $\Ham_{\textrm{dis}}$ 
and $\Ham_{\textrm{hop}}$, representing, respectively, free motion along
each edge, impurity scattering and interlayer hopping, 
and a contribution $\Ham_{\textrm{int}}$ from Coulomb interactions.
We write it in terms of the electron creation operator $c\dagg_{qn}$
for an edge state with wavevector $q$ in layer $n$, taking sample
perimeter $L$ so that $q=2\pi n_q/L$, where $n_q$ is integer.
The creation operator at a point is
\begin{equation}
\psi_n^{\dagger}(x)=\frac{1}{\sqrt{L}}\sum_{q=-\infty}^{\infty}
e^{-i q x }c_{qn}^{\dagger}\label{eq:ctopsi}\,.
\end{equation}
We normal order the Hamiltonian with respect to a vacuum in
which states are occupied for $q\leq 0$ and empty otherwise.
Then
\begin{eqnarray}
\Ham_0&=&
-i\hbar v\sum_n\int dx
:\!\psi^{\dagger}_n(x)\partial_x\psi_n(x)\!:\label{eq:H0psi}\,,
\end{eqnarray}
and
\begin{align}
\Ham_{\textrm{hop}}&=
\sum_n\int dx [t_{\perp}\psi^{\dagger}_{n+1}(x)
\psi_n(x)+\textrm{H.~c.}]\,.
\end{align}
The interaction contribution, written in terms of the projected density 
$\rho(x)=\psi^{\dagger}_n(x) \psi_n(x)$ with a two-particle potential
$U_{n-m}(x-x')$, is
\begin{equation}
\Ham_{\textrm{int}}
=\frac{1}{2}\sum_{nm}\int dx \int dx' :\rho_n(x) U_{n-m}(x-x')\rho_m(x'):\,.
\end{equation}

Finally, writing the impurity potential projected
onto the edge coordinate in the $n$th layer as $V_n(x)$,
we have
\begin{equation}
\Ham_{\textrm{dis}}=\sum_n\int dx
V_n(x):\psi^{\dagger}_n(x)\psi_n(x):
\, .
\end{equation}
We take $V_n(x)$ to be Gaussian distributed with
zero-range correlations and strength $\Delta$: 
$[V_n(x)]_{\textrm{av}}=0$ and
$[V_n(x)V_{n'}(x')]_{\textrm{av}}=\Delta\delta_{n,n'}\delta(x-x')$.
This disorder term can be removed by
means of a gauge transformation on the fermionic field
operators, under which
\begin{equation}
\psi_n^{\dagger}(x)\to
e^{i\theta_n(x)}\psi_n^{\dagger}(x)\label{eq:gaugexform},
\end{equation}
where
\begin{equation}
\theta_n(x)=\frac{1}{\hbar v}\int_0^xdx'V_n(x')
\end{equation}
is the phase shift acquired under forward scattering
from the impurities. The elastic scattering length is
related to the disorder strength $\Delta$ by $\lel=\hbar^2
v^2/\Delta$.  Under this gauge transformation,
$\Ham_0+\Ham\subtxt{dis}\to\Ham_0$. The hopping term, however, picks
up a dependence on the disorder, and after the transformation
is 
\begin{equation}
\Ham_{\textrm{hop}}=\sum_n\int dx
[\tperp(n,x)\psi^{\dagger}_{n+1}(x)
\psi_n(x)+\textrm{H.~c.}]\label{eq:Hhop},
\end{equation}
where
\begin{equation}
\tperp(n,x)=\tperp
e^{i(\theta_{n+1}(x)-\theta_n(x))}\label{eq:hopphase}.
\end{equation}
We ignore the effects of this gauge transformation
on the boundary conditions applying to $\psi_n(x)$,
which is justified at temperatures large compared to the 
single-particle level spacing.
With this, $\Ham_{\textrm{0}}+\Ham_{\textrm{int}}$ is unaffected
by the gauge transformation, and gauge transformed operators
$c_{qn}^{\dagger}$ can be defined by inverting Eq.~(\ref{eq:ctopsi}). 
All further references in this
paper to fermionic operators are to the gauge-transformed ones.

%
%
\subsection{Bosonised Hamiltonian}\label{ssec:model:bose}

We bosonize the Hamiltonian in the standard way, expressing
$\Ham_{\textrm{0}}+\Ham_{\textrm{int}}$ in terms of non-interacting
collective modes. Since $\Ham_{\textrm{hop}}$ transforms
into a cosine function of the boson creation and annihilation operators,
we treat it perturbatively. To justify this, we require that $t_\perp$
should be small. Since $t_\perp$ is a relevant perturbation,\cite{Naud1}
we also require that temperature should not be too small: 
$L_{\rm T} \ll l_\perp$.

Boson creation operators are defined in the
usual way (see, for example, Ref.~\onlinecite{vonDelft}) as
\begin{equation}
b_{qm}^{\dagger}=\frac{i}{(n_q)^{1/2}}\sum_{r=-\infty}^{\infty}
c^{\dagger}_{r+q,m} c\phdagg_{r,m}
\end{equation}
for $q>0$.  
Fourier transforming the interaction potential and expressing the 
result as a velocity, we introduce
\begin{equation}
u_{n-m}(q) = (2\pi \hbar)^{-1}\int dx e^{iqx}U_{n-m}(x)\,.
\end{equation}
The Fermi velocity renormalised by Hartree interactions is
$v_{\rm F} = v - \sum_n u_n(0)$, where the divergence 
which arises in the sum in the case of Coulomb interactions
is cancelled by contributions to $v$ from a neutralising background.
The Hamiltonian in the absence of hopping (and omitting
fermion number terms which appear at electron densities different from
that of our vacuum) is
\begin{equation}
\Ham_0+\Ham_{\textrm{int}}=\sum_{mn}\sum_{q>0}\hbar[v_{\rm F} + 
u_{n-m}(q)]q b\dagg_{qn}b\phdagg_{qm}\,.
\end{equation}

The combination $\Ham_0+\Ham\subtxt{int}$ is diagonalised by Fourier
transform in the layer index $n$. We impose periodic boundary conditions
on $n$, define the wavevector $k=2n_k\pi/Na$, with $n_k$ integer and
$-\pi/a \leq k < \pi/a$, and set
\begin{equation}
b^{\dagger}_{qk}=\frac{1}{\sqrt{N}}\sum_{n=1}^N
e^{inka}b^{\dagger}_{qn}\,,
\end{equation}
and
\begin{equation}
u(q,k) = \sum_n e^{inka} u_n(q)\,.
\end{equation}
Then
\begin{equation}
\Ham_0+\Ham_{\textrm{int}}=\sum_k
\sum_{q>0}\hbar\omega(q,k)b^{\dagger}_{qk}b\phdagg_{qk}\label{eq:bosonH}
\end{equation}
where the excitation frequencies are 
\begin{equation}
\omega(q,k)=[\vF+u(q,k)]q\label{eq:defomega}.
\end{equation}

The Coulomb interaction, regularised at short distances by a finite width
$w$ for edge states, has the form
\begin{equation}
U_n(x)=\frac{e^2}{4\pi\epsilon_0\epsilon_r}\frac{1}{\sqrt{x^2 +n^2a^2 + w^2}}\label{eq:rsCoulomb}\,.
\end{equation}
The edge state width $w$ is set by the localisation length $\xi$
of localised states in the bulk of the sample at the Fermi energy.
In a clean sample with well-separated Landau levels, $\xi \sim l_{\rm B}$,
but in a highly disordered sample with Landau levels that are broad in energy
one may have $\xi \gg l_{\rm B}$. The value of $w$ proves important
in matching our results to experiment, as we discuss in Sec.~\ref{ssec:cond:results}.

We write the Fourier transform, using the Poisson summation formula, as
\begin{equation}
u(q,k)=v_F\frac{\kappa }{2\pi} \sum_p \iint
dxdz\frac{e^{-i(qx+kz+2\pi p z/a)}}{\sqrt{x^2+z^2+w^2}}\,.
\end{equation}
and find
\begin{equation}
\omega(q,k)=\vF
q\left(1+\kappa\sum_{p\,\in\mathbb{Z}}Q_p^{-1}e^{-wQ_p}\right)\label{eq:CoulombFT}
\end{equation}
with $Q_p^2=q^2+(k+2\pi p/a)^2$ and $p$ integer. Here, the inverse screening length
$\kappa\equiv e^2/4\pi\epsilon\subtxt{r}\epsilon_0 \hbar \vF a$ characterises
the interaction strength.

For isolated layers, taking the limit of large $a$, the sum on $p$
may be replaced with an integral and one recovers the
dispersion relation of edge magnetoplasmons in 
a single layer system, known from previous work.\cite{Volkov1,Volkov2} 

For the
multilayer system the
expression for the dispersion relation may be simplified in two stages. First,
if the layer spacing is small ($a\ll w$) the sum on $p$ may be
omitted, so that
\begin{equation}
\omega(q,k)=\vF q\left(1+\frac{\kappa
e^{-w\sqrt{q^2+k^2}}}{\sqrt{q^2+k^2}}\right)\label{eq:dispwide}\,.
\end{equation}
If, in addition, interactions are weak ($w \ll \kappa^{-1}$)
\begin{equation}
\omega(q,k)=\vF
q\left(1+\frac{\kappa}{\sqrt{q^2+k^2}}\right)\label{eq:dispnarrow}.
\end{equation}
In the following we obtain detailed results for systems
with wide edges using the dispersion relation of Eq.~(\ref{eq:dispwide}),
and for systems with narrow edges using the dispersion
relation of Eq.~(\ref{eq:dispnarrow}).

%
%
\subsection{Two-particle correlation function}\label{ssec:model:G}

A central quantity in our calculations of transport properties is the
two-fermion correlation function
\begin{equation}
G(x,t)\equiv\langle\psi_n^{\dagger}(x,t)\psi\phdagg_{n+1}(x,t)
\psi^{\dagger}_{n+1}(0,0)\psi\phdagg_n(0,0)\rangle\label{eq:defG}\,,
\end{equation}
where $\langle\ldots\rangle\equiv\textrm{Tr}(e^{-\beta
\Ham}\ldots)/\textrm{Tr}(e^{-\beta \Ham})$ and
operators are written in the Heisenberg representation, with $\mathcal{O}(t)=e^{i\Ham
t/\hbar}\mathcal{O}e^{-i\Ham t/\hbar}$.
We evaluate this in the absence of tunneling, so that
$\Ham=\Ham_{0}+\Ham_{\textrm{int}}$. 

As a first step, define the boson field operator\cite{footnote0}
\begin{equation}
\phi_n(x)=-\sum_{q>0}n_q^{-1/2}\:
\left(e^{-iqx}b_{qn}^{\dagger}+ e^{iqx}b_{qn}\right)e^{-\epsilon
q/2}\label{eq:defphiboson}
\end{equation}
where $\epsilon$ is a short-distance cut-off.
Omitting Klein factors (which cancel from
$G(x,t)$), the fermion and boson field operators are related by
\begin{equation}
\psi_n(x)=(2\pi\epsilon)^{-1/2}\exp{(-i\phi_n(x))}\,.
\end{equation}
The correlation function is
\begin{equation}
G(x,t)\!\!=\!\frac{1}{(2\pi\epsilon)^2}\langle
e^{i\phi_n(x,t)}e^{-i\phi_{n+1}(x,t)}e^{i\phi_{n+1}(0,0)}e^{-i\phi_n(0,0)}\rangle.
\end{equation}
We define its logarithm $S$ via
\begin{equation}
G(x,t)\equiv\frac{1}{(2\pi)^2}e^S\label{eq:GtoS}\,.
\end{equation}

Because $\Ham$ is harmonic, $S$ can be expressed as
\begin{equation}
\begin{split}
S=&-\frac{1}{2}\left<(\phi_n(x,t)\!-\!\phi_{n+1}(x,t)\!+\!\phi_{n+1}(0,0)\!-\!\phi_n(0,0))^2\right>\\
&+\!\frac{1}{2}\left[\phi_n(x,t)\!-\!\phi_{n+1}(x,t),\phi_n(0,0)\!-\!\phi_{n+1}(0,0)\right]\\
&-2\log{\epsilon}.
\end{split}
\end{equation}
The thermal average and the commutator appearing in this expression can be
evaluated in the standard way via a mode expansion, by expressing
$\phi_n(x,t)$ in terms of boson creation and annihilation
operators using Eq.~(\ref{eq:defphiboson}).
Taking the thermodynamic limit and so
replacing wavevector sums with integrals, with $\beta = 1/k_{\rm B} T$,
we arrive at
\begin{align}
&S(x,t,T)=-2\log{\epsilon}-\frac{a}{\pi}\!\int^{\pi/a}_{-\pi/a}\!\!\!\!\!\!dk(1-\cos{ak})
\!\!\int_0^{\infty}\!\!\frac{dq}{q}e^{-\epsilon q}\nonumber\\
&\!\times\!\!\bigg(\!\!\coth{\!\left(\,\beta\hbar\omega(q,k)/\,2\,\right)}\!
\left[1\!-\!\cos{(qx\!-\omega(q,k)t)}\right]\label{eq:fullS}\\
&\qquad\qquad\qquad\qquad\qquad\qquad+i\sin{(qx\!-\omega(q,k)t)}
\!\!\bigg)\nonumber.
\end{align}
It is useful to note that
\begin{equation}
G(-x,-t)=G(x,t)^{\ast}\label{eq:Gstar},
\end{equation}
and also to define a frequency-dependent correlator,
\begin{equation}
\tilde{G}(x,\Omega)=\int dt e^{i\Omega t}G(x,t).
\end{equation}

\section{Conductivity}\label{sec:cond}

In this section we express the conductivity $\sigma(T)$
obtained from a Kubo formula in terms of the two-fermion 
correlation function calculated in Sec.~\ref{ssec:model:G}.
We also set out the steps required for a numerical evaluation
of $\sigma(T)$, present our results, and compare them with the experimental
data of Ref.~\onlinecite{UCSB10}. 

\subsection{Kubo formula for conductivity}\label{ssec:cond:kubo}

The operator for the interlayer current density between layers $n$
and $n+1$ is
\begin{equation}
j_n(x)=\frac{ie}{\hbar}
\left(\tperp(n,x)\psi_{n+1}^{\dagger}(x)\psi_n(x)-\textrm{H.~c.}\right).\label{eq:defj}
\end{equation}
The real part of the conductivity at frequency $\Omega$ 
is given by the Kubo formula\cite{footnote}
\begin{align}
\sigma(\Omega,T)\!=\!\frac{ia}{\hbar\Omega
L}\sum_m\!\int_{-\infty}^{\infty} \!\!&dt\sin{\Omega t}\!\!\int
\!dx\!\int\! dx'\nonumber\\
&\times\left\langle\!
j_n(x,t)j_m(x'\!,0)\!\right\rangle\label{eq:sigmajj}\, .
\end{align}
To leading order, the interlayer hopping appears only in the current
operators, and we evaluate the thermal average 
using a Hamiltonian from which interlayer hopping is omitted.

Substituting for $j_n(x,t)$ using
Eq.~(\ref{eq:defj}) gives an expression for the conductivity of
the chiral metal with a given configuration of disorder: to
leading order in $t_{\perp}(n,x)$,
\begin{align}
\sigma(\Omega,T)&=\frac{2iaL}{\hbar\Omega}\left(\frac{e}{\hbar
L}\right)^2\!\!\int\! dx\!\int\!
dx'\!\int_{-\infty}^{\infty}\!\!\!\!
dt\sin{\Omega t}\label{eq:sigmadis}\nonumber\\
&\times\tperp(n,x)\tperp^\ast(n,x')\nonumber\\
&\times\langle\psi_n^{\dagger}(x,t)\psi_{n+1}(x,t)\psi^{\dagger}_{n+1}(x',0)\psi_n(x',0)\rangle\,.
\end{align}
Averaging over disorder configurations yields
\begin{equation}
[\tperp(n,x)\tperp^\ast(n,x')]_{\rm av} = \tperp^2 e^{-|x|/l_{\rm el}}
\end{equation}
and hence
\begin{align}
\sigma(\Omega,T)&=\frac{e^2}{h}\frac{8\pi i a \lel
t_{\perp}^2}{\Omega\hbar^2}\int \frac{dx}{2\lel}\, e^{-\vert
x\vert/\lel}\!\int_{-\infty}^{\infty}\!dt \sin{\Omega t}\nonumber\\
&\times\langle\psi_n^{\dagger}(x,t)\psi_{n+1}(x,t)\psi^{\dagger}_{n+1}(0,0)\psi_n(0,0)\rangle\label{eq:sigmaGfull}.
\end{align}
This result can be expressed in terms of the time or frequency
dependent two-particle correlation functions defined in
Sec.~\ref{ssec:model:G}. Setting $\Omega=0$ we find
\begin{align}\label{sigma}
\sigma(T)&\!=-\frac{e^2}{h}\frac{8\pi a \lel
t_{\perp}^2}{\hbar^2}\!\!\int\!\!\frac{dx}{2\lel}\, e^{-\vert
x\vert/\lel}\!\!\!\int_{-\infty}^{\infty}\!\!\!\!\!\!dt\,t\,\textrm{Im}G(x,t)\\
&\equiv\frac{e^2}{h}\frac{8\pi a \lel
t_{\perp}^2}{\hbar^2}\!\!\int\!\!\frac{dx}{2\lel}\, e^{-\vert
x\vert/\lel}\textrm{Re}\left[\partial_{\,\Omega}
\tilde{G}(x,\Omega)\big\vert_{\Omega=0}\right]\nonumber.
\end{align}

For a boson dispersion relation $\omega(q,k)= v_{\rm F}q$, as results
from the Hartree approximation, the fermion correlation function factorises
into independent contributions from each layer. These
have the form
\begin{equation}
\langle\psi_n^{\dagger}(x,t)\psi_{n}(0,0)\rangle=\frac{1}{2\pi}\int_{-\infty}^{\infty}dk
\frac{e^{ik(\vF t-x)}}{1+e^{\,\beta\hbar\vF k}}
\end{equation}
and we find a temperature-independent conductivity
\begin{equation}
\sigma(\Omega,T)=\frac{e^2}{h}\frac{2t_{\perp}^2\lel
a}{\hbar^2\vF^2}\frac{1}{1+\Omega^2\lel^2/\vF^2}\,,
\end{equation}
which in the zero-frequency limit has the value
\begin{equation}
\sigma_0 =\frac{e^2}{h}\frac{2t_{\perp}^2l_{el}a}{\hbar^2\vF
^2}\label{eq:sigma0}\,.
\end{equation}

More generally, with an arbitrary boson dispersion relation a simplification
of Eq.~(\ref{sigma}) is possible for $\lel\ll\LT$, since $G(x,t)$
varies with $x$ only on the scale $\LT$ while the
correlator $[\tperp(n,x)\tperp^\ast(n,x')]_{\rm av}$ has range $\lel$.
We get
\begin{align}
\sigma(T)&=-4\pi\sigma_0\vF^2\int_{-\infty}^{\infty}\!\!\!dt\,t\,
\textrm{Im}\,G(0,t)\label{eq:sigmaGIm}\nonumber\\
&\equiv\phantom{-}4\pi\sigma_0
\vF^2\textrm{Re}\left[\partial_{\,\Omega}\tilde{G}(0,\Omega)
\big\vert_{\Omega=0}\right].
\end{align}
%
%
\subsection{Evaluation of $\sigma(T)$}\label{ssec:cond:evalsigma}

To find the temperature dependence of the conductivity we must
combine Eqs. (\ref{eq:GtoS}), (\ref{eq:fullS}), and
(\ref{eq:sigmaGIm}). A first step before numerical evaluation
is to isolate the dependence on the cut-off
$\epsilon$ and take the limit $\epsilon \rightarrow 0$,
as we describe in this subsection.

We start from the expression given in Eq.~(\ref{eq:fullS})
for the
logarithm of the two-particle correlation function,
which we evaluate at $x=0$. It is convenient to separate
out a zero-temperature contribution by writing
\begin{equation}
S(t,T)\equiv S(t,0)+\Delta S(t,T)
\end{equation}
and also to split $S(t,0)$ into real and imaginary parts, with
\begin{equation}
S(t,0)\equiv \mathcal{U}(t)-i\mathcal{V}(t)\,,
\end{equation}
where $\mathcal{U}(t)$ and $\mathcal{V}(t)$ are real for $t$ real.  
Then, 
writing
\begin{equation}
\sigma(T)=\sigma(0)+\Delta\sigma(T),
\end{equation}
we obtain from
Eq.~(\ref{eq:sigmaGIm})
\begin{align}
\sigma(0)=&\frac{2\sigma_0
\vF^2}{\pi}\int_0^{\infty}\!\!\!dt\,t
\,e^{\mathcal{U}(t)}\sin{\mathcal{V}(t)}\label{eq:sigma0UV}
\end{align}
and
\begin{align}
\Delta\sigma(T)\!\!=&\frac{2\sigma_0\vF^2}{\pi}\!\!
\int_0^{\infty}\!\!\!\!\!dt\,t\,e^{\mathcal{U}(t)}\sin{\mathcal{V}(t)}\!
\left[e^{\Delta S(t,T)}\!-1\right]\label{eq:DeltasigmaUV}.
\end{align}
In the case of a linear boson dispersion relation, 
$\omega(q,k)= v_{\rm F}q$, the functions $\U(t)$ and $\V(t)$
have the forms
\begin{align}
\mathcal{U}_{\,\textrm{lin}}(t)&=-\log{(\epsilon^2+\vF^2t^2)}\label{eq:defUlin}\\
\mathcal{V}_{\,\textrm{lin}}(t)&=\pi-2\tan^{-1}(\epsilon/\vF
t)\label{eq:defVlin}.
\end{align}
Adding and subtracting these expressions from the ones for $\U(t)$
and $\V(t)$ with a general dispersion relation, we find
\begin{align}
&\mathcal{U}(t)=\mathcal{U}_{\,\textrm{lin}}(t)+\frac{a}{\pi}\int_{-\pi/a}^{\pi/a}\!\!\!dk(1-\cos{ak})\nonumber\\
&\times\int_0^{\infty}\frac{dq}{q}\,e^{-\epsilon
q}[\cos{(\omega(q,k)t})-\cos{(\vF qt)}]\label{eq:defU}
\end{align}
and
\begin{align}
&\mathcal{V}(t)=\mathcal{V}_{\,\textrm{lin}}(t)+\frac{a}{\pi}\int_{-\pi/a}^{\pi/a}\!\!\!dk(1-\cos{ak})\nonumber\\
&\times\int_0^{\infty}\frac{dq}{q}\,e^{-\epsilon q}
\left[\sin{(\omega(q,k)t)}-\sin{(\vF qt)}\right]\label{eq:defV}.
\end{align}
Finally, we have
\begin{align}
\Delta S(t,T)&=-\frac{a}{\pi}\int_{-\pi/a}^{\pi/a}\!\!\!dk(1-\cos{ak})\label{eq:defDeltaS}\\
\times\int_0^{\infty}\frac{dq}{q}&e^{-\epsilon
q}(1-\cos{\omega(q,k)t})
\left[\coth{\left(\frac{\beta\hbar\omega(q,k)}{2}\right)}-1\right]\nonumber.
\end{align}

The advantage of casting the equations for the conductivity in
this form is that the momentum integrals in Eqs.~(\ref{eq:defU}),
(\ref{eq:defV}) and (\ref{eq:defDeltaS}) can be performed at $\epsilon=0$, since the
integrands decay fast enough at large $q$ for convergence.  
Dependence on $\epsilon$ is confined for small $\epsilon$ 
to the functions $\Ulin(t)$ and $\Vlin(t)$, and from
Eqs.~(\ref{eq:defUlin}) and (\ref{eq:defVlin}) one sees 
that it is important only for
$t\sim\order{\epsilon}$.
It is therefore convenient to separate
the integration range in 
Eq.~(\ref{eq:sigma0UV}) into two parts, $0\leq t<R$ and
$R\leq t <\infty$, with $\epsilon\ll R\ll 1$.
In the first interval $\mathcal{U}(t)=\Ulin(t)$ and $\mathcal{V}=\Vlin(t)$;
in the second interval one can set $\epsilon=0$.

Let the contributions to $\sigma(0)$ from the two intervals be
$\sigma^{(1)}$ and
$\sigma^{(2)}$.  Writing 
$t'=\vF t/\epsilon$ we have
\begin{equation}
\sigma^{(1)}=\frac{2\epsilon^2\sigma_0}{\pi}\int_0^{R\epsilon^{-1}}\!\!\!\!\!dt'\,t'\,
e^{\mathcal{U}(t')}\sin{\mathcal{V}(t')}
\end{equation}
which gives
\begin{equation}
\sigma^{(1)}=\frac{2\sigma_0}{\pi}\int_{0}^{\infty}dt'\,t'\,
\frac{1}{1+t'^2}\frac{2t'}{1+t'^2}=\sigma_0\,.
\end{equation}
Evaluation of $\sigma^{(2)}$ requires a numerical calculation, 
and we present results in Sec.~\ref{ssec:cond:results}.

Finally, turning to the conductivity at non-zero temperature,
we note that there are no extra difficulties
in the evaluation of $\Delta\sigma$
using Eq.~(\ref{eq:DeltasigmaUV}).
The function $\Delta S(t,T)$,
can be computed
numerically with $\epsilon=0$, and $\Delta S(t,T)\rightarrow 0$
as $t \rightarrow 0$, so that $\Delta\sigma(T)$ has no contribution
from the integration interval $0\leq t<R$ in the limit
$\epsilon \rightarrow 0$.

In summary, when evaluating $\sigma(0)$ or $\Delta\sigma(T)$ using
Eqs.~(\ref{eq:sigma0UV}) and (\ref{eq:DeltasigmaUV}), the functions
$\mathcal{U}(t)$, $\mathcal{V}(t)$, and $\Delta S(t,T)$ may be
evaluated numerically by setting $\epsilon=0$ in
Eqs.~(\ref{eq:defU}), (\ref{eq:defV}), and (\ref{eq:defDeltaS}),
and the results used in Eq.~(\ref{eq:sigma0UV})
to find $\sigma^{(2)}$. To this one must add
$\sigma^{(1)}=\sigma_0$ in order to obtain the zero temperature
conductivity $\sigma(0)$.  These equations combine
with Eq.~(\ref{eq:DeltasigmaUV}) for $\Delta\sigma(T)$ to give a
computationally tractable, though non-trivial, expression for $\sigma(T)$.
%
%
\subsection{Conductivity at zero temperature}\label{ssec:cond:sigma0T}

The conductivity at zero temperature and zero frequency is determined
solely by the low energy limit of the group velocity for excitations,
since no other modes are excited as $T,\Omega \rightarrow 0$.
This zero frequency limit is reached as $q$, the wavevector component
in the chiral direction, approaches zero. The
group velocity, $\partial \omega(q,k) / \partial q\vert_{q=0} \equiv \vF \alpha(k)$,
is in general a function of $k$, the wavevector component in
the interlayer direction.

To determine $\sigma(0)$, a useful procedure is to consider a model dispersion 
relation which is exactly linear
in $q$: $\omega(q,k)=\vF
q\alpha(k)$. A linear dispersion relation is also of interest in its own right.
It arises from an interaction that in real space is short range
in the chiral direction, $x$: $U_n(x)=g_n\delta(x)$, giving
$\alpha(k)=1+(2\pi\hbar\vF)^{-1}\sum_{n}e^{ikna}g_n$.
With a linear dispersion relation, $q$-integrals in
the expressions leading to $G(x,t)$ can be evaluated analytically,
greatly simplifying the calculation of conductivity. 
As we show in the following, for the limit $l_{\rm el} \ll L_{\rm T}$ that we consider,
a dispersion relation linear in $q$ yields a temperature-independent
value of conductivity.
For interactions, such as the Coulomb potential, that are not short
range in $x$, linearisation of the dispersion relation gives only an approximation
to $G(x,t)$. The value of $\sigma(0)$ that results 
from integrating this approximate form for $G(x,t)$
is nevertheless exact (at the leading order in $t_{\perp}$
considered throughout this paper). 
This fact is clear on physical grounds, since
we have correctly accounted for the dispersion relation at low energy.
It may also be derived formally, as follows.

Starting from Eq.~(\ref{eq:sigmaGIm}),
we deform the contour for the time integral 
into the semicircle at infinity in the lower half of the complex plane,
writing $t=t_R + {\rm i} t_I$ with $t_R$ and $t_I$ real.
Then in Eq.~(\ref{eq:fullS}) we have the factor
\begin{equation}
\int_0^{\infty} dq \frac{1}{q} \exp(-\epsilon q -iqx -it_R\omega(q,k) + t_I \omega(q,k))\,.
\end{equation}
This must be evaluated for all values of $t$ lying on the deformed time integration contour.
When $|t_R|$ is large, $\exp(-it_R\omega(q,k))$ is a rapidly oscillating function of $q$,
and the $q$-integral can be computed using the method of stationary phase: since
$\omega(q,k)$ is a monotonically increasing function of $q$, the dominant contribution
comes from the vicinity of the end-point at $q=0$. Similarly, when $t_I$ is large and negative,
$\exp(t_I\omega(q,k))$ is small for most values of $q$, and
the $q$-integral can be computed using steepest descents: again, the dominant contribution
comes from the vicinity of $q=0$. In both instances we may approximate
$\omega(q,k)$ by its form linearised about $q=0$; after linearisation
the $q$-integral can be evaluated analytically.

This calculation yields
\begin{equation}
\begin{split}
G(0,t)=\frac{1}{(2\pi)^2}\left(\frac{\pi t/\beta\hbar}{\sinh{(\pi
t/\beta\hbar)}}\right)^2\frac{1}{\vF^2}\frac{1}{(\epsilon+it)^2}\\
\times\exp{\left(-\frac{2a}{\pi}\int_0^{\pi/a}\!\!\!dk\,(1-\cos{ak})\log{\alpha(k)}\right)}.
\end{split}
\end{equation}
Substituting this into Eq.~(\ref{eq:sigmaGIm}) we obtain
\begin{align}
\sigma(T)=\,&\frac{2\sigma_0}{\pi}\exp{\left(-\frac{2a}{\pi}
\int_0^{\pi/a}\!\!\!dk\,(1-\cos{ak})\log{\alpha(k)}\right)}\nonumber\\
&\times\int\frac{dt\,\epsilon
t^2}{(\epsilon^2+t^2)^2}\left(\frac{\pi t/\beta\hbar}{\sinh{(\pi
t/\beta\hbar)}}\right)^2.
\end{align}
In the limit $\epsilon\to 0$, the $t$ integral gives $\pi/2$
regardless of temperature, demonstrating that, for systems
with a linear dispersion relation, in the regime $l_{\rm el} \ll L_{\rm T}$,
$\sigma(T)$ is independent of $T$.  We find
\begin{equation}
\sigma(T)=\sigma_0\exp{\left(-\frac{2a}{\pi}\int_0^{\pi/a}
\!\!\!dk(1-\cos{ak})\log{\alpha(k)}\right)}\label{eq:sigmaT=0}\,.
\end{equation}
This is our final result for the dependence of $\sigma(0)$
on the dispersion relation as parameterised by $\alpha(k)$.
%
%
\subsection{Results}\label{ssec:cond:results}

We are now in a position to calculate the conductivity
for a system with Coulomb interactions by
evaluating numerically the formulae we have derived: first, the
zero-temperature value using the results from Sec.~\ref{ssec:cond:sigma0T},
and then the full temperature-dependent conductivity using the
results from Sec.~\ref{ssec:cond:evalsigma}.  We investigate variation
of the conductivity with two parameters, the Fermi velocity $\vF$ and
the edge state depth $w$, and seek values of these parameters for
which our results match the experimental data of Ref.~\onlinecite{UCSB10}. 
The parameters enter
the dispersion relation $\omega(q,k)$ directly, and $\vF$ also
appears in the inverse screening length $\kappa$.
The interaction strength is set by the combination $\kappa a$
(recall that $a$ is the layer spacing).
A scale for temperature is set by $\vF$ and $a$, via
$T_0\equiv \hbar\vF/a\kB$, so that $T/T_0=a/\LT$.
A scale for conductivity
is given by $\sigma_0$, its value in the Hartree approximation.

At a qualitative level, the effect of interactions on the conductivity
can be anticipated by starting from the expression
given in Eq.~(\ref{eq:sigma0}) for this quantity
within the Hartree approximation. In turn, that expression
can be understood in terms of a calculation
of the interlayer tunneling rate, based on the Fermi golden rule:
the rate involves the square of a matrix element between initial and
final states on adjacent layers, and a power of the density of states
for both the initial and the final states. The squared matrix element, allowing for
disorder which affects phases of initial and final states separately,
contributes a factor of $t_\perp^2 \lel$ to $\sigma_0$. The
form of the density of states on a single edge, $1/2\pi \hbar \vF$, 
implies that $\sigma_0 \propto \vF^{-2}$. Returning to
a full treatment of the interacting system, we note that the 
effect of interactions is to generate an energy-dependent group
velocity in place of a constant value, $\vF$. 
In effect, the value of $\sigma(T)$ at a particular temperature
involves a thermal average of the inverse square of the group velocity.
Because Coulomb interactions increase the group velocity at low
energy, they decrease conductivity at low temperature;
equally, because the group velocity approaches $\vF$ at high energy,
the conductivity approaches $\sigma_0$ at high temperature.

\begin{figure}
\begin{center}
\includegraphics[width=0.40\textwidth]{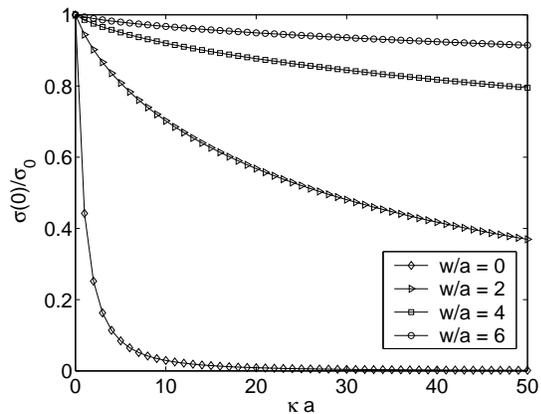}
\caption{\label{fig:sigmaT0} Conductivity at zero temperature, as a function
of interaction strength, parameterised by 
inverse screening length $\kappa$, for various edge state widths $w$.}
\end{center}
\end{figure}

Turning to detailed results,
the dependence of $\sigma(0)$ on
$w/a$ and $\kappa a$ is shown in Fig.~\ref{fig:sigmaT0},
as obtained from Eq.~(\ref{eq:sigmaT=0}) using $\alpha(k) = 1 +\kappa e^{-w|k|}/|k|$.  
Interactions reduce the value of the conductivity, by a factor
which is large if $\kappa a$ is large.
The variation of $\sigma(T)$ with $T$ is illustrated in
Fig.~\ref{fig:sigmaTw=0kappa=1,5} for a system with
the dispersion relation appropriate for
narrow edge states,
Eq.~(\ref{eq:dispnarrow}). In this case
the $k$ integrals
in Eqs.~(\ref{eq:defU}) and (\ref{eq:defV}) can be done
analytically, leaving only the $q$ and $t$ integrals to 
be evaluated numerically.
\begin{figure}
\begin{center}
\includegraphics[width=0.40\textwidth]{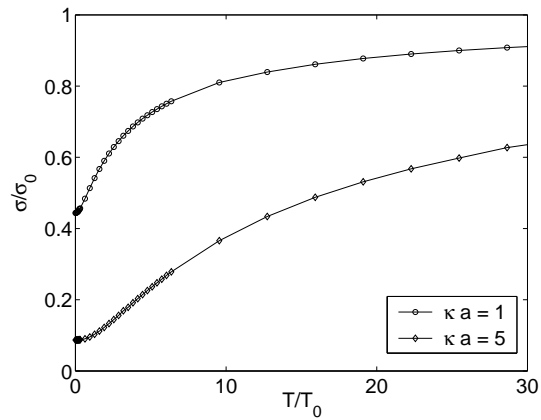}
\caption{\label{fig:sigmaTw=0kappa=1,5} Dependence of conductivity on
temperature for narrow edge states, with interaction strengths $\kappa a=1$ and
$\kappa a=5$. }
\end{center}
\end{figure}
Finally, the behaviour of $\sigma(T)$ for a system with
wide edge states ($w\geq a$) is presented in Fig.~\ref{fig:sigmaTw=4kappa=1,50}.
In this case the dispersion relation is as given in Eq.~(\ref{eq:dispwide}), 
analytical progress does not seem possible, 
and integrals on $k$, $q$ and $t$ must be evaluated numerically to obtain
$\sigma(T)$.
\begin{figure}
\begin{center}
\includegraphics[width=0.40\textwidth]{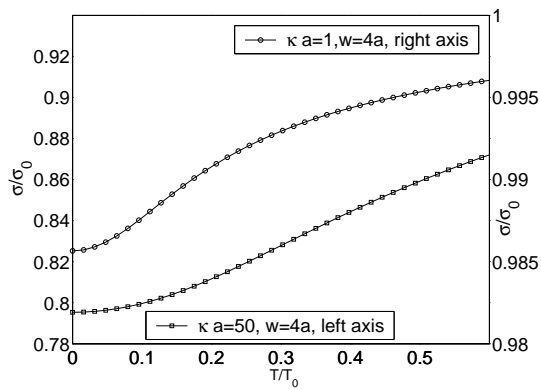}
\caption{
\label{fig:sigmaTw=4kappa=1,50}
Dependence of conductivity on
temperature for wide edge states 
with $w=4a$ and interaction strengths 
$\kappa
a=1$ 
and 
$\kappa 
a=50$.}
\end{center}
\end{figure}
We note in passing that we checked that there are only
small changes to the results presented when using the more complete form
of the interaction given in Eq.~(\ref{eq:CoulombFT}), including the sum on $p$.

Examining these results, it is evident that the general shape of $\sigma(T)$
does not vary greatly with parameters: the temperature
dependence is quadratic at low temperatures, has a roughly
linear region at intermediate temperatures, and approaches $\sigma_0$ 
in the high temperature limit.
The quadratic dependence at low temperature is universal,
but the extent of the roughly linear region at intermediate
temperature is model-dependent. Moreover,
scales in this temperature dependence change dramatically
with parameter values. The value of the dimensionless temperature
$T/T_0$ at the crossover between the low  and intermediate
temperature regimes is dependent on $\kappa$ (see
Fig.~\ref{fig:sigmaTw=0kappa=1,5}) and varies even more strongly
with $w$ (compare Figs. \ref{fig:sigmaTw=0kappa=1,5} and
\ref{fig:sigmaTw=4kappa=1,50}). In addition, the magnitude of
the variation in $\sigma(T)$ between low and high $T$ depends
very much on the values of $w$ and $\kappa a$.
In order to reproduce the experimental observation
of a nearly linear increase in $\sigma(T)$, by about 7\% between the 
temperatures of 50mK and 300mK,\cite{UCSB10} we require parameters which
place the experimental temperature window in the intermediate regime
for behaviour, so that quadratic variation of $\sigma(T)$ with $T$
occurs only in a temperature range below 50 mK, and saturation of $\sigma(T)$
occurs only above 300mK.
Since the available data is not sufficiently detailed
to justify a formal fitting procedure, we instead survey
the consequences of a range of parameter choices in our results
and examine the match to experimental observations.

We begin by considering
narrow edges states, using the results shown in
Fig.~\ref{fig:sigmaTw=0kappa=1,5}. Supposing $\vF \sim \omega_{\rm C} l_{\rm B}$,
which represents an upper bound on $\vF$, 
we have $\vF = 1.7 \times 10^5{\rm ms}^{-1}$.
With $a=30$nm, we find $\kappa a \sim 1$ 
and $T_0 \sim 40$K. Taking these values, the variation in $\sigma(T)$ 
over the experimental temperature range is
very small and quadratic, in disagreement with observations.
A reduction in the value of $\vF$ serves to decrease the temperature scale $T_0$,
and also increases $\kappa$. 
It is possible to generate approximately linear variation of $\sigma(T)$
with $T$ in the experimental temperature range by using a sufficiently small
value of $\vF$ (reduced from the upper bound by $\sim\order{10^3}$), 
but we know of no reason for $\vF$ to be so small.

\begin{figure}
\begin{center}
\includegraphics[width=0.40\textwidth]{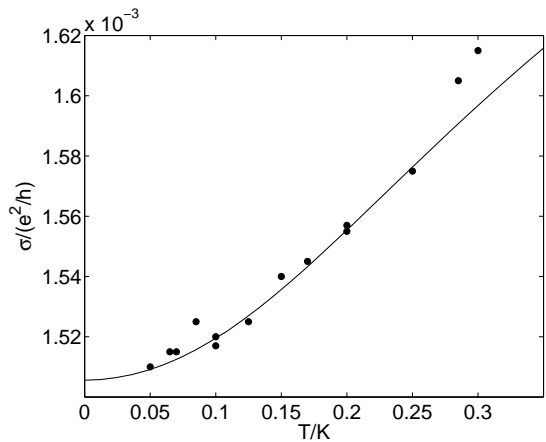}
\caption{\label{fig:sigma(T)w=4kappa=50}Dependence of conductivity
on temperature for
$w=4a$, with $a=30\textrm{nm}$, 
$\vF =3\times10^{3}\textrm{ms}^{-1}$ and
$\sigma_0=1.893 \times 10^{-3}e^2/2\pi \hbar$ (full line),
compared with experimental data (points) taken from Fig 2 
of Ref.~\onlinecite{UCSB10} (data set for Fractal 2).
}
\end{center}
\end{figure}

We therefore turn to theoretical results for wide edge states,
as illustrated in Fig.~\ref{fig:sigmaTw=4kappa=1,50}.
In this case, we find that large values of $w$ 
greatly reduce the temperature range over which $\sigma(T)$ varies
quadratically with $T$, and can
lead to approximately linear variation in the experimental
temperature range. A second consequence of large $w$ is that
the conductivity change $\sigma(\infty) - \sigma(0)$ is reduced.
This tendency can be counteracted by increasing the 
interaction strength $\kappa a$.
We find that observed behaviour can be reproduced
by taking $w=4a=120{\rm nm}$ and
$\vF=3\times10^3\textrm{ms}^{-1}$ (giving $\kappa a=50$).  The temperature dependence
of $\sigma(T)$ obtained using these parameter values is
shown in Fig.~\ref{fig:sigma(T)w=4kappa=50} for temperatures
below 400mK.

This choice of parameters, and its implications, merit further discussion.
First, we note that there are two separate experimental indications
that edge states have a width closer to the value we have adopted, of
$120 {\rm nm}$, than to the conventionally expected value of 
$l_{\rm B} \simeq 10{\rm nm}$. 
One comes from measurements of bulk hopping transport
in multilayer samples\cite{UCSB13}, 
which give a localisation length of
$\xi = 120{\rm nm}$: one expects $w\simeq \xi$. The other comes
from studies of
conductance fluctuations,\cite{UCSB12} 
discussed in Sec.~\ref{sec:condflucs}.
These yield a value for the inelastic scattering length,
from the amplitude of fluctuations, and a value for
the area of a phase-coherent region perpendicular to the
applied field, from the correlation field for 
fluctuations. The ratio of this phase-coherent area to the inelastic scattering length
implies an edge state width which is also much larger than
$l_{\rm B}$: $w\simeq 70{\rm nm}$.
Next, turning to the value of $\vF$, which we have taken $50$
times smaller than for edge states in a steep confining potential,
we note that large edge state width favours a small value for
$\vF$, because wide edge states penetrate into the bulk of the sample
where both the confining potential gradient and the drift velocity
of electrons moving in this potential are small. 
Finally, we comment on the fact that accepting a small value for
$\vF$ implies a large value for $\sigma_0$, 
if other parameters are unchanged. In fact, large $w$ acts in
the opposite direction, to reduce the effective tunneling amplitude
$t_\perp$ between edge states, since different portions of the edge
contribute to the amplitude with different phases, so that there are
partial cancellations. To account for the magnitude of the measured\cite{UCSB10}
conductivity, $1.5\times 10^{-3} e^2/2\pi \hbar$, using
the value for the mean free path $\lel = 30 {\rm nm}$ derived from
magnetoresistance measurements\cite{UCSB11} requires an effective value of $t_\perp$
about 50 times smaller than bare estimate\cite{UCSB1} of $0.12$ meV.
This is a surprisingly strong supression of tunneling, though possible if edge states
in successive layers have different displacements from the surface, as suggested in
Ref.~\onlinecite{UCSB10}.


\section{Conductance fluctuations}\label{sec:condflucs}

It is found experimentally that mesoscopic fluctuations
in the conductance of the chiral metal are induced
by small changes of magnetic field within a quantum Hall 
plateau.\cite{UCSB9,UCSB12} These conductance fluctuations are
observed in samples with a perimeter that is several times larger
than the estimated inelastic scattering length.
Under such conditions, it is not initially clear why the
magnetic field component perpendicular to layers in the sample 
should influence conductance in this way, since
in the simplest picture electron trajectories 
enclose flux only by encircling the sample. More realistically,
a number of possibilities are evident:\cite{UCSB12}
the sample walls may lie at an angle to the layer normal,
either on average or because of surface roughness,
or finite edge state width may be important.
In our theoretical treatment of conductance fluctuations we
avoid specific assumptions about this aspect of the system
by considering fluctuations that result from
variations in a magnetic field component $B_\perp$ perpendicular to
the sample surface. The amplitude of fluctuations
is not affected by this choice. By contrast, the scale for the
correlation field of fluctuations is
dependent on the model chosen for flux linkage.

In a general setting, there are two possible
reasons for the amplitude of conductance fluctuations
to decrease with inceasing temperature. One is because of a decrease in
the inelastic scattering length; the other is 
because of thermal smearing. In the case of a chiral metal
only the first mechanism operates, because states at
different energies are perfectly correlated.\cite{Betouras}
In this sense, conductance fluctuations offer a
rather direct probe of interaction effects.

In this section, in place of conductivity $\sigma$, 
we are concerned with the conductance
$g=\sigma L/Na$ of a finite sample and fluctuations
$\delta g = g - [g]_{\rm av}$ about its average value.
We denote the average  within the Hartree approximation
by $g_0\equiv \sigma_0 L/Na$.
We derive an analytic expression for the autocorrelation
function of conductance fluctuations induced by $B_\perp$.
We focus on its temperature dependence at low temperatures,
obtaining a scaling form for the
regime in which $\sigma(T) \approx \sigma(0)$.
We compute the scaling function, evaluate our expressions numerically, 
and compare our results with the observations of Ref.~\onlinecite{UCSB12}.

%
%
\subsection{Correlation function}\label{ssec:condflucs:corfn}

The conductance autocorrelation function

\begin{equation}
F(\delta B)=[\delta g(B_{\perp}) \delta g(B_{\perp}+\delta
B)]_{\textrm{av}}
\end{equation}
is characterised by the amplitude $F(0)$
and by the correlation field. An obvious field scale is set by a
flux density of one flux quantum $\Phi_0$ through a rectangle with sides
proportional to the layer spacing and the thermal length, and we define
$B_0=\Phi_0/2\pi a\LT =\hbar/eaL_{\rm T}$. We also introduce a dimensionless
field variation $b=\delta B/B_0$, which depends on temperature through 
$L_{\rm T}$, and a temperature-independent reduced field $h$
which has dimensions of wavevector:
$h=b/L_{\rm T} \equiv e \delta B/a\hbar$.

With a suitable choice of gauge, the transverse field enters
the Hamiltonian only as a phase for interlayer hopping.
Taking for convenience $B_\perp =0$, in the
presence of non-zero $\delta B$
Eq.~(\ref{eq:hopphase}) is modified to
\begin{align}
\tperp(n,x)=\tperp e^{i(\theta_{n+1}(x)-\theta_n(x)+hx)}.
\end{align}
This additional, field-dependent phase alters $\Ham\subtxt{hop}$
and consequently the current operator.

An expression for the conductance of a sample with a specific
disorder configuration is obtained by scaling
Eq.~(\ref{eq:sigmadis}) with the sample dimensions.
Taking account of the field-dependent phases in the current operator and
substituting into the definition of $F(\delta B)$, after some
manipulation we arrive at 
\begin{align}
F(\delta B)&=\frac{g_0^2\pi^2\vF^4}{L^2\lel^2N^2}\sum_{n,m}\int
dx\int dx'\int dy\int dy'\label{eq:flucsfull}\\
\times&\int dt\,it\,G(x-x',t)\int
dt'\,it'\,G(y-y',t')\nonumber\\
\times&\left(e^{ih(x-x')}+e^{-ih(x-x')}\right)e^{C(x,x')}e^{C(y,y')}\nonumber\\
\times&
\left(e^{D_{nm}(x,x';\,y,y')}+e^{-D_{nm}(x,x';\,y,y')}-2\right)\nonumber\,.
\end{align}
Two contributions to this expression arise from the disorder
average:
\begin{equation}
C(x,x')\!=-\frac{1}{2}[(\theta_{n+1}(x)\!-\!\theta_n(x)
\!-\!\theta_{n+1}(x')\!+\!\theta_n(x'))^2]_{\textrm{av}}
\end{equation}
and
\begin{align}
D_{nm}&(x,x';y,y')\!=\!
\big[(\theta_{n+1}(x)\!-\!\theta_n(x)\!-\!\theta_{n+1}(x')\!+\!\theta_n(x'))\nonumber\\
\times&(\theta_{m+1}(y)\!-\!\theta_m(y)\!-\!\theta_{m+1}(y')\!+\!\theta_m(y'))\big]_{\textrm{av}}.
\end{align}
Both may be evaluated using the result (for $x,\,y>0$)
\begin{equation}
[\theta_n(x) \theta_m(y)]_{\textrm{av}}=\frac{\delta_{nm}}{\lel}\min\{x,y\}\,.
\end{equation}
The equation for $C$ gives
\begin{equation}
e^{C(x,x')}=e^{-|x-x'|/\lel},
\end{equation}
which in the limit of small $\lel$ can be written $2\lel\delta(x-x')$.
The expression for $D$ is more complicated: one finds
\begin{align}
D_{nm}(x,x';y,y')\!=\!\frac{R(x,x';y,y')}{\lel}
\left(2\delta_{nm}\!\!-\!\delta_{n+1,m}\!\!-\!\delta_{n-1,m}\right).
\end{align}
The function $R(x,x';y,y')$ gives the overlap between the two
directed intervals on the real line $x\to x'$ and $y\to y'$:
for example, $R(1,5;4,9)=-R(5,1;4,9)=1$. On substituting these
expressions for $C$ and $D$ into Eq.(\ref{eq:flucsfull}), we
obtain
\begin{align}
&F(b)=\frac{g_0^2\pi^2\vF^4}{L^2\lel^2N}\!\!\int\!\!
dx\!\!\int\!\! dx'\!\!\int\!\! dy\!\!\int\!\! dy'\big(e^{ih(x-x')}+e^{-ih(x-x')}\big)\nonumber\\
&\times\int dt\,it\,G(x-x',t)\int
dt'\,it'\,G(y-y',t')e^{-|x-x'|/\lel}\nonumber\\
&\times
e^{-|y-y'|/\lel}\bigg\{e^{2R(x,x';\,y,y')/\lel}+e^{-2R(x,x';\,y,y')/\lel}-2\nonumber\\
&+\,2e^{R(x,x';\,y,y')/\lel}+2e^{-R(x,x';\,y,y
')/\lel}-4\bigg\}\label{eq:huge}.
\end{align}

Examining where the weight of the integrand lies with respect to
the spatial integrals in Eq.~(\ref{eq:huge}), one sees that the
term in braces vanishes except in places where $R\ne0$.  We
consider different types of contributions from these regions, and
keep only those which are leading order for $\LT \gg\lel$.  First,
consider regions in which $\vert x-y\vert\sim \lel$ but $\vert
x-x'\vert\gg \lel$. The small factor
$e^{-|x-x'|/\lel}$ is compensated by
the first term in the braces if $\vert x'-y'\vert\sim\lel$. Then
\begin{align}
e^{-|x-x'|/\lel}e^{-|y-y'|/\lel}e&^{2R(x,x';\,y,y')/\lel}
=\\&\qquad e^{(-|x-y|-|x'-y'|)/\lel}\nonumber.
\end{align}
Since $G(x,t)$ has a range in $x$ of order $\LT$, the
resulting contribution to $F(\delta B)$ is $\order{\LT/L}$.  Another
contribution of the same order arises from regions where $\vert
x-y'\vert\sim\lel$ and $\vert x'-y\vert\sim\lel$. Subleading
contributions come from regions where all four spatial variables
are within an elastic length of one another. These contributions
are $\order{\lel/L}$.

Keeping only the leading order terms, the 
expression for the correlation function has the much simplified form
\begin{align}
F(\delta B)&=\frac{4g_0^2\pi^2\vF^4}{NL}\int dx (e^{ihx}+e^{-ihx})\int
\!\!dt\,it\,\int\!\! dt'\,it'\nonumber\\
&\times\left(G(x,t)G(x,t')+G(x,t)G(-x,t')\right).
\end{align}
Using the symmetry of $G(x,t)$ (see Eq.~(\ref{eq:Gstar})) one finds
\begin{equation}
F(\delta B)=\frac{g_0^2}{NL}\,\int_{-\infty}^{\infty}\!\!dx\,
e^{ihx}[f(x)]^2\label{eq:cfcorr}\,,
\end{equation}
where
\begin{equation}
f(x)\equiv
-4\pi\vF^2\int_{-\infty}^{\infty}\!\!dt\,t\,\textrm{Im}G(x,t)\label{eq:deff}\,.
\end{equation}
%
%
\subsection{Computing the correlation function}\label{ssec:condflucs:computing}

In order to compare our theory for conductance fluctuations with
experiment, we need to be able to calculate $F(\delta B)$ for
various values of the temperature and parameters $\vF$ and $w$.
Although it is possible to use a computer to evaluate the form of
$F(B_{\perp})$ given in Eq.~(\ref{eq:cfcorr}) without further
approximation, it is far easier to make progress by calculating
$G(x,t)$ for a linearised
dispersion relation. 
This
approach is exact in the low-temperature regime defined by the condition
$\sigma(T)\approx \sigma(0)$, and we proceed to use it
in our calculations.

In the low temperature regime where the linearised dispersion
relation may be used, $F(B_{\perp})$ has a scaling form.
To make this apparent, it is helpful to recast equations in terms of dimensionless
variables,
characterising $\delta B$ by $b$ in place of $h$,
and introducing $\hat{x}=x/L_T$ and $\hat{t}= \vF t/\LT$.
Writing $G(x,t)=(2\pi \LT)^{-2} \hat{G}(\hat{x},\hat{t})$ and
$f(\LT \hat{x})=\hat{f}(\hat{x})$, for
a linear dispersion
relation, $\omega(q,k)=q\vF\alpha(k)$, we have
\begin{align}
&\hat{G}(\hat{x},\hat{t})=\exp\bigg\{\frac{-2a}{\pi}\int_0^{\pi/a}
\!\!\!\!\!dk(1-\cos{ak})\nonumber\\
&\times\bigg[\log|\hat{x}-\alpha(k)
\hat{t}|\nonumber-\log\bigg(\frac{\pi[\alpha(k)
\hat{t}-\hat{x}]/\alpha(k)}{\sinh(\pi[\alpha(k)
\hat{t}-\hat{x}]/\alpha(k))}\bigg)\bigg]\bigg\}\\
&\times\exp\bigg\{\!\!-ia\!\!\int_0^{\pi/a}
\!\!\!\!\!dk(1-\cos{ak})\,\textrm{sgn}\,(\hat{x}-\alpha(k)
\hat{t})\bigg\}\label{G-hat}
\end{align}
and
\begin{equation}
\hat{f}(\hat{x})=-\frac{1}{\pi}\int_{-\infty}^{\infty}d\hat{t}\,\hat{t}\,
\textrm{Im}\{\hat{G}(\hat{x},\hat{t})\}\label{eq:fredefine}\,.
\end{equation}
Then the conductance autocorrelation function has the form
\begin{equation}
F(\delta B)=\frac{g_0^2\LT}{NL}\,C\left(\delta
B/{B_0}\right)\label{eq:corrfinal}
\end{equation}
with scaling function
\begin{equation}\label{C(b)}
C(b)=\int_{-\infty}^{\infty}
d\hat{x}\,e^{ib\hat{x}}[\hat{f}(\hat{x})]^2.
\end{equation}
In this form $F(\delta B)$ depends on temperature $T$ and magnetic field difference $\delta B$
only through the scaling variables $\LT/L$ and $\delta B/B_0$. 
The thermal length $\LT$ plays
the role of an inelastic scattering length, in the sense that it determines both the amplitude
of conductance fluctuations and (through $B_0$) their correlation field.
Such behaviour is initally surprising, since $\LT$ is independent of interaction strength.
In fact, of course, the form of the scaling function $C(b)$
depends parametrically on interaction strength.

For weak interactions this dependence of $C(b)$ on $\kappa$ can be extracted analytically,
as follows. First, note from Eq.~(\ref{eq:dispnarrow}) that $\alpha(k) = 1+\kappa/|k|$.
Also, in Eqs.~(\ref{G-hat}), (\ref{eq:fredefine}) and (\ref{C(b)}), change variables from
$\hat{x},\hat{t}$ to $y,p$ with $\hat{x}=y/\kappa$ and $\hat{t}=yp+y/\kappa$.
Then
\begin{align}
\lim_{\kappa\rightarrow 0}\,&\hat{G}(y/\kappa,p+y/\kappa)\equiv g(y,p)\nonumber\\ 
=\,&\exp\bigg\{\frac{-2a}{\pi}\int_0^{\pi/a}
\!\!\!\!\!dk(1-\cos{ak})\nonumber\\
&\times\bigg[\log|y(p+1/k)|
-\log\bigg(\frac{\pi y[p+1/k]}
{\sinh(\pi y[p+1/k])}\bigg)\bigg]\bigg\}\nonumber\\
&\times\exp\bigg\{ia\!\!\int_0^{\pi/a}
\!\!\!\!\!dk(1-\cos{ak})\,\textrm{sgn}\,(y[p+1/k])\bigg\}\nonumber
\end{align}
and
\begin{equation}
\lim_{\kappa\rightarrow 0} \,\hat{f}(y/\kappa ) \equiv \tilde{f}(y) = 
-\frac{y^2}{\pi}\int_{-\infty}^{\infty}dp\,p\
\textrm{Im}\{{g}(y,p)\}\,.\nonumber
\end{equation}
The $\kappa$-dependence of the scaling function is hence isolated for small $\kappa$
as
\begin{equation}
C(b) = \frac{1}{\kappa}\int_{-\infty}^{\infty} dy \exp(iyb/\kappa) [\tilde{f}(y)]^2\,,
\end{equation}
demonstrating that the amplitude of conductance fluctuations
grows and that the correlation field shrinks as interactions are made weaker.
In both cases, the variation implies an inelastic scattering length that diverges
as $\kappa^{-1}$ for weak interactions.
Such a dependence of the inelastic scattering length on interaction strength
is long-established in non-chiral, one-dimensional conductors.\cite{Apel}

In order to find the form of the scaling function and to study its 
$\kappa$-dependence at general $\kappa$, 
a three-dimensional numerical integration is necessary.
We compute $\hat{G}(\hat{x},\hat{t})$, then
$\hat{f}(\hat{x})$, and then the scaling function $C(b)$ itself. 
%
%
\subsection{Results}\label{ssec:condflucs:results}

\begin{figure}
\begin{center}
\includegraphics[width=0.40\textwidth]{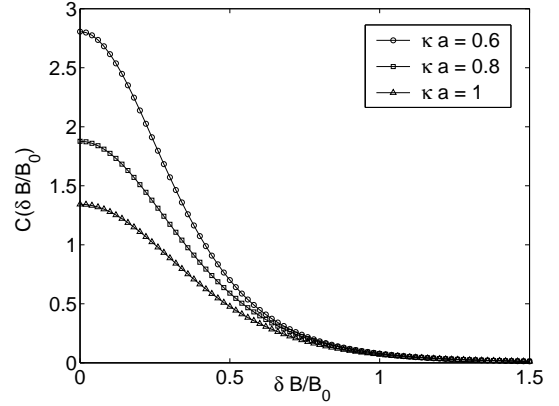}
\caption{\label{fig:flucsw=0} $C(\delta B/B_0)$ for narrow edge states and
$\kappa a=0.6$, $0.8$, and $1$. }
\end{center}
\end{figure}

\begin{figure}
\begin{center}
\includegraphics[width=0.40\textwidth]{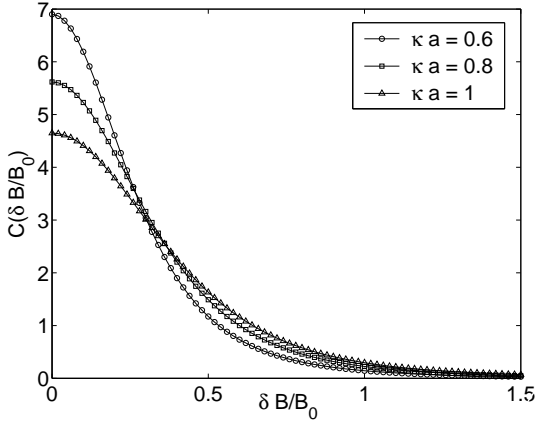}
\caption{\label{fig:flucsw=1} $C(\delta B/B_0)$ for $w=a$ and
$\kappa a=0.6$, $0.8$, and $1$. }
\end{center}
\end{figure}

\begin{figure}
\begin{center}
\includegraphics[width=0.385\textwidth]{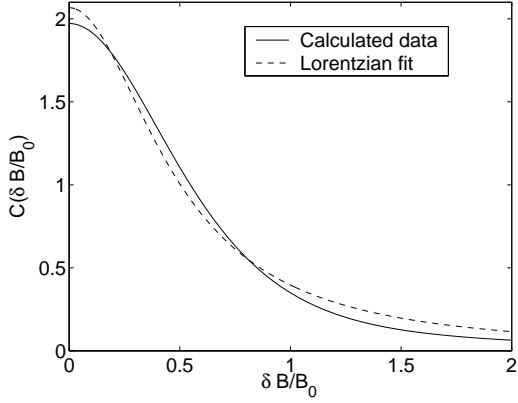}
\caption{\label{fig:Fscaling} $C(\delta B/B_0)$ at $w=4a$ and
$\kappa a=50$.}
\end{center}
\end{figure}

\begin{figure}
\begin{center}
\includegraphics[width=0.40\textwidth]{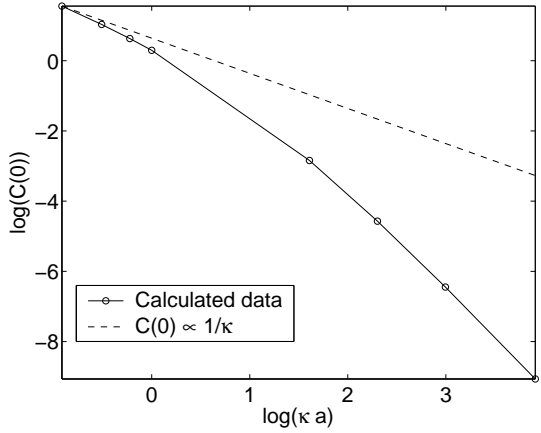}
\caption{\label{fig:flucskappa} Conductance fluctuation amplitude
as a function of interaction strength $\kappa a$ at $w=0$ (full line),
and asymptotic behaviour calculated analytically for small $\kappa a$
(dashed line)}
\end{center}
\end{figure}

We illustrate the form of the
scaling function $C(\delta B/ B_0)$  for a range of parameter
values in a sequence of three figures. Its dependence on interaction strength
$\kappa a$ is shown for narrow edge states in Fig.~\ref{fig:flucsw=0}
and for $w=a$ in Fig.~\ref{fig:flucsw=1}. In both cases, smaller
interaction strength leads to a larger amplitude for conductance fluctuations
and a smaller correlation field, as may be anticipated on the grounds that
weaker interactions lead to a longer inelastic scattering length.
In Fig.~\ref{fig:Fscaling} $C(\delta B/B_0)$ is shown  for
$\kappa=50$ and $w=4a$, the
parameter values suggested by the comparison of our conductivity
calculations with experiment. We discuss experimental data on 
conductance fluctuations in Sec.~\ref{ssec:condflucs:expt}.
Finally, the increase in the amplitude of conductance fluctuations
with dereasing $\kappa$ is illustrated in 
Fig.~\ref{fig:flucskappa}. 
%
%
\subsection{Comparison with experiment and previous
theory}\label{ssec:condflucs:expt}

\begin{figure}
\begin{center}
\includegraphics[width=0.40\textwidth]{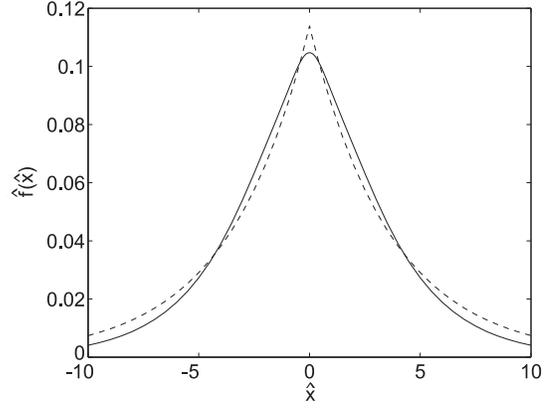}
\caption{\label{fig:f} $\hat{f}(\hat{x})$ calculated at $\kappa a=50$
and $w=4a$ (solid line) compared with the best fitting exponential
decay (dashed line).}
\end{center}
\end{figure}

The exact treatment of disorder and interactions
provided by the calculations we have decribed presents an opportunity
to test the standard theoretical treatment of conductance fluctuations,
in which a single inelastic scattering length $l_{\rm in}$,
or equivalently a scattering rate $\vF/l_{\rm in}$ is used as a cut-off
in perturbation theory. For the chiral metal, such calculations have been
described in Ref.~\onlinecite{Betouras}. They 
yield a Lorentzian scaling function
\begin{equation}
F(\delta
B)=\frac{2g_0^2}{NL}\frac{l\subtxt{in}}{1+z^2}\label{eq:CBcorrelator}
\end{equation}
with $z=2\pi \delta Bl\subtxt{in}a/\Phi_0$.
A comparison between the functional form we obtain for 
$F(\delta B)$ and a Lorenztian is given in
Fig.~\ref{fig:Fscaling}: while the two functions are similar,
the discrepancies are worth attention because they
indicate behaviour which cannot be characterised by a single
relaxation time. A similar comparison can be made in
the Fourier transformed domain, in terms of the function $f(x)$.
To reproduce Eq.~(\ref{eq:CBcorrelator}) from our Eq.~(\ref{eq:cfcorr}),
we would require $l\subtxt{in}=\LT$ and
\begin{equation}
\hat{f}(\hat{x})= e^{-\vert \hat{x}\vert /2}\,,
\end{equation}
where exponential decay is indicative of a single lifetime
$l\subtxt{in}/\vF$ for excitations. The form we obtain
for $\hat{f}(\hat{x})$ is shown in Fig.~\ref{fig:f}.
The absence of a cusp at $x=0$ indicates that there is 
of a range of relaxation times in the system.
In addition, the fact that $f(0)\not= 1$ is an interaction
effect (from Eq.~(\ref{eq:sigmaGIm}) one sees that $f(0)=\sigma(0)/\sigma_0$)
not allowed for in the standard perturbative treatment.

We close this section with a comparison between the experiments
of Ref.~\onlinecite{UCSB12} and our results, using the same parameters,
$\kappa a=50$ and $w=4a$, that provided a match for the behaviour of $\sigma(T)$.
For the experimental base temperature of
$T=70\textrm{mK}$, we use our approach to determine the amplitude 
of conductance fluctuations. As a way to present the result,
we then follow the experimental analysis\cite{UCSB12} in using 
Eq.~(\ref{eq:CBcorrelator}) to obtain a value for $\lin$ of
$0.3\mu\textrm{m}$.  The experimental value,
extracted in the same way, is
$\lin\sim 1\mu\textrm{m}$.  
Since the calculated amplitude of conductance fluctuations
varies by several orders of magnitude over the range of
parameter values we have investigated, and since no
new adjustment of parameters was involved in our
discussion of conductance fluctuations,
we find the rough agreement between these two values of $\lin$ very encouraging.

\section{Conclusions}\label{sec:discussion}

In summary, for the system of weakly coupled
quantum Hall edge states that we have studied, bosonisation
provides a very complete treatment of the interplay between
electron-electron interactions and
disorder. We have shown that interaction effects
can account for the observed temperature dependence
of interlayer conductivity, provided we allow for
finite edge state width and adopt a value for the edge state velocity
that is rather smaller than previously supposed.
We have investigated conductance fluctuations within
the same theoretical approach, showing how they are
suppressed with increasing temperature, with a characteristic lengthscale
$\LT \propto T^{-1}$. Encouragingly,
the same parameter values used to match
the measured behaviour of conductivity 
reproduce approximately the observed 
fluctuation amplitude.
From a theoretical
viewpoint, it is interesting that such dephasing effects
can be generated from a description based on harmonic
collective modes, simply via the nonlinear
relation between boson and fermion operators.

\begin{acknowledgments}

We thank E. G. Gwinn for very helpful discussions and J. J.
Betouras for previous collaborations.  The work was supported by
EPSRC under Grant GR/R83712/01 and by the Dutch FOM foundation.

\end{acknowledgments}

\bibliographystyle{h-physrev}
\bibliography{bibliography}

\begin{thebibliography}{10}
\bibitem{footnote0} 
Eqns.~(\ref{eq:defphiboson}) and (\ref{eq:fullS}) correct sign errors
in Eqns.~(8) and (13) of Ref.~\onlinecite{PRL}.
None of the results of Ref.~\onlinecite{PRL} are
changed by these corrections.
\end{thebibliography}

\end{document}